\documentclass[12pt,reqno]{amsart}
\usepackage{amsmath,amssymb,amsfonts,amsthm}
\usepackage[mathscr]{eucal}
\usepackage[all]{xy}
\usepackage{hyperref}
\usepackage{setspace}
\usepackage{bm}
\usepackage{xcolor}
\allowdisplaybreaks

\author{V.~A.~Abakumova, S.~L.~Lyakhovich}

\address{Physics Faculty, Tomsk State University, Lenin ave. 36, Tomsk 634050, Russia}

\email{abakumova@phys.tsu.ru, \, sll@phys.tsu.ru}

\title{Global conserved quantities and unfree gauge symmetry}

\begin{document}

\maketitle

\begin{abstract}
We consider a class of theories with unfree gauge symmetry, whose gauge parameters are restricted by differential equations. We demonstrate that such theories admit global conserved quantities, whose on-shell values are defined by asymptotics of the fields rather than Cauchy data. The global conserved quantities can be deduced proceeding from the equations restricting gauge parameters, and they are treated differently by two BRST complexes corresponding to a system with unfree gauge symmetry.
\end{abstract}
\section{Introduction}
\noindent Let us consider the Lagrangian equations, $\partial_iS(\phi)=0$, subject to the modified Noether identities
\begin{equation}\label{LNI}
    \partial_iS(\phi)\Gamma^i{}_\alpha
    -\tau_a(\phi)\Gamma^a{}_\alpha\equiv0\,,
\end{equation}
which result in unfree gauge symmetry transformations \cite{Kaparulin:2019quz}
\begin{equation}\label{Lunfree}
    \delta_\epsilon\phi^i=\Gamma^i{}_\alpha
    \epsilon^\alpha\,; \quad \delta_\epsilon S(\phi)=0\,\,
    \Rightarrow\,\, \Gamma^a{}_\alpha\epsilon^\alpha=0\,,
\end{equation}
if $\Gamma^a{}_\alpha$ is a differential operator with finite kernel, $u_a\Gamma^a{}_\alpha=0$. The quantities
\begin{equation}\label{Ltau}
     \mathcal{T}_a(\phi,\Lambda)\equiv\tau_a(\phi)
     -u_a(\Lambda)\approx 0\,, \quad u_a(\Lambda)=\Lambda_{\mathcal{I}}
     u^{\mathcal{I}}{}_a\,,
\end{equation}
which do not reduce to the linear combination of Lagrangian equations, contain constants $\Lambda_\mathcal{I}$ treated as modular parameters. The simplest example of a modular parameter is the cosmological constant of unimodular gravity. Its analogues for higher spin field theories with transverse symmetry \cite{Skvortsov:2007kz}, \cite{Campoleoni:2012th} are found in \cite{Abakumova:2020ajc}, \cite{Abakumova:2021mun}.
The completion functions (\ref{Ltau}) and their differential consequences can be resolved with respect to the modular parameters, $\Lambda_{\mathcal{I}}\approx J_{\mathcal{I}}(\phi)$, so there exist global conserved quantities $J_\mathcal{I}$, whose specific values are defined by asymptotics of the fields and their derivatives.

Any action with unfree gauge symmetry admits the alternative reducible symmetry with unconstrained gauge parameters \cite{Francia:2013sca}. The global conserved quantities are treated differently depending on the type of symmetry. In the next Section, we demonstrate it in constrained Hamiltonian formalism. 

\section{Global conserved quantities in constrained Hamiltonian formalism}

\noindent Let us consider the constrained Hamiltonian system
\begin{equation}\label{SH}
    S_H=\int dt\big(p_i\dot{q}^i-H(q,p)-\lambda^\alpha T_\alpha(q,p)\big)\,,
\end{equation}
with involution relations
\begin{equation}
    \{T_\alpha, H\}=\tau_a \Gamma^a{}_\alpha\,, \quad \{\tau_a\,, H\}=T_\alpha V^\alpha{}_a\,,
\end{equation}
where $\Gamma\neq\Gamma(q,p), V\neq V(q,p)$, and all the constraints commute. The kernel of differential operator $\Gamma^a{}_\alpha$ is supposed to be finite. The irreducible secondary constraints $\tau_a$ contain modular parameters, $\tau_a=\tau_a(q,p,\Lambda)$. The unfree gauge symmetry transformations read
\begin{equation}\label{unfreetransf}
    \delta_\epsilon q^i=\{q^i, T_\alpha\}\epsilon^\alpha
    +\{q^i,\tau_a\}\epsilon^a\,, \quad \delta \lambda^\alpha=\dot{\epsilon}^\alpha
    +V^\alpha{}_a\epsilon^a\,, \quad \dot{\epsilon}^a
    +\Gamma^a{}_\alpha\epsilon^\alpha=0\,.
\end{equation}
For more detailed description of constrained Hamiltonian formalism for the systems with unfree gauge symmetry, see 
\cite{Abakumova:2019uoo}, \cite{Abakumova:2021rlq}. 

The complete BRST charge reads,
\begin{equation}\label{Q}
    Q=T_\alpha C^\alpha+\tau_aC^a+\pi_\alpha P^\alpha\,,
\end{equation}
where $T_\alpha$ and $\tau_a$ are primary and secondary constraints, respectively, and all the ghosts and Lagrange multipliers are introduced in Table \ref{table:unfree}. 

\begin{table}[h]
\caption{Ghosts and Lagrange multipliers
(unfree gauge symmetry)}\label{table:unfree}
\begin{center}
\begin{tabular}{| c | c | c | c | c || c | c | c | c | c |}
\hline
\phantom{@}$q$\phantom{@} & \phantom{@}$p$\phantom{@} & gh $q$  & gh $p$ & \phantom{@}$\varepsilon$\phantom{@} &\phantom{@}$q$\phantom{@} & \phantom{@}$p$\phantom{@} & gh $q$  & gh $p$ & \phantom{@}$\varepsilon$\phantom{@}\\ \hline
$C^{\alpha^{\phantom'}}$ & $\overline{P}_\alpha$ & 1  & -\,1 & 1 &$\lambda^{\alpha^{\phantom'}}$ & $\pi_\alpha$ & 0  & 0 & 0\\ \hline
$C^{a^{\phantom'}}$ & $\overline{P}_a$ & 1  & -\,1 & 1 &$P^{\alpha^{\phantom'}}$ & $\overline{C}_\alpha$ & 1  & -\,1 & 1\\ \hline
\end{tabular}
\end{center}
\end{table}

If we consider $\widetilde{\tau}_\alpha=\tau_a\Gamma^a{}_\alpha$ as reducible secondary constraints of the second stage of reducibility, i.e. there exist null-vectors $Z_1{}^\alpha{}_A$ and $Z_2{}^A{}_{A_1}$, such that the involution relations read
\begin{equation}
    \{T_\alpha, H\}=\widetilde{\tau}_\alpha\,, \quad 
    \{\widetilde{\tau}_\alpha,H\}=
    T_\beta\widetilde{V}^\beta{}_\alpha\,, \quad
    \widetilde{\tau}_\alpha Z_{1}{}^\alpha{}_A=0\,, \quad
    Z_1{}^\alpha{}_AZ_2{}^A{}_{A_1}=0\,,
\end{equation}
where $\widetilde{V}\neq\widetilde{V}(q,p)$, $Z_1\neq Z_1(q,p)$, $Z_2\neq Z_2(q,p)$. The corresponding gauge symmetry transformations will be reducible,
\begin{equation}\label{redtransf}
    \begin{array}{c}
   \delta_\varepsilon q^i=
    \{q^i,T_\alpha\}\big(\dot{\varepsilon}^\alpha
    -Z_1{}^\alpha{}_A\varepsilon^A\big)
    -\{q^i,\widetilde{\tau}_\alpha\}\varepsilon^\alpha\,, \\[2mm]
    \delta_\varepsilon \lambda^\alpha=
    \big(\ddot{\varepsilon}^\alpha
    -Z_1{}^\alpha{}_A\dot{\varepsilon}^A\big)-
    \widetilde{V}^\alpha{}_\beta\varepsilon^\beta\,;\\[2mm]
    \delta_\omega\varepsilon^\alpha=
    Z_1{}^\alpha{}_A\omega^A\,, \quad \delta_\omega\varepsilon^A=\dot{\omega}^A
    -Z_2{}^A{}_{A_1}{\omega}^{A_1}\,;\\[2mm]
    \delta_\eta\omega^A=Z_2{}^A{}_{A_1}\eta^{A_1}\,,
    \quad \delta_\eta\omega^{A_1}=\dot{\eta}^{A_1}\,.
    \end{array}
\end{equation}
The corresponding BRST charge reads,
\begin{equation}\label{widetildeQ}
    \begin{array}{c}
    \widetilde{Q}=T_\alpha C^\alpha
    +\widetilde{\tau}_\alpha\mathcal{C}^\alpha
    +\overline{\mathcal{P}}_\alpha Z_1{}^\alpha{}_A\mathcal{C}^A
    +\overline{\mathcal{P}}_A
    Z_2{}^A{}_{A_1}\mathcal{C}^{A_1}\\[2mm]
    +\,\pi_\alpha P^\alpha
    +\pi_A\mathcal{P}^A+\pi_{A_1}\mathcal{P}^{A_1}
    +\pi'_{A_1}\mathcal{P}'^{A_1}\,,
    \end{array}
\end{equation}
all the ghosts and Lagrange multipliers are introduced in Table {\ref{table:reducible}}. For more details, see \cite{Abakumova:2021mun}.

So, the same Hamiltonian system (\ref{SH}) admits two different BRST complexes. They are related, as $Q$ (\ref{Q}) and $\widetilde{Q}$ (\ref{widetildeQ}) are connected by some non-local canonical transformation \cite{Abakumova:2021mun}, but non-equivalent. Indeed, the irreducible secondary constraints $\tau_a$, being the Hamiltonian counterparts of the completion functions $\mathcal{T}_a$ (\ref{Ltau}), are BRST-exact with respect to $Q$ (\ref{Q}), and BRST-closed, but not BRST-exact with respect to $\widetilde{Q}$ (\ref{widetildeQ}). Hence, the modular parameters are treated as trivial quantities in the case of unfree gauge symmetry, and as physical observables in a theory with reducible gauge symmetry. In the next Section, we demonstrate this inequivalence in the model of Maxwell-like spin 2 model.

\begin{table}[h]
\caption{Ghosts and Lagrange multipliers
(reducible gauge symmetry)}\label{table:reducible}
\begin{center}
\begin{tabular}{| c | c | c | c | c || c | c | c | c | c |}
\hline
\phantom{@}$q$\phantom{@} & \phantom{@}$p$\phantom{@} & gh $q$  & gh $p$ & \phantom{@}$\varepsilon$\phantom{@} &\phantom{@}$q$\phantom{@} & \phantom{@}$p$\phantom{@} & gh $q$  & gh $p$ & \phantom{@}$\varepsilon$\phantom{@}\\ \hline
$C^{\alpha^{\phantom'}}$ & $\overline{P}_\alpha$ & 1  & -\,1 & 1 &$\lambda^{\alpha^{\phantom'}}$ & $\pi_\alpha$ & 0  & 0 & 0\\ \hline
$\mathcal{C}^{\alpha^{\phantom'}}$ & $\overline{\mathcal{P}}_\alpha$ & 1  & -\,1 & 1 &$P^{\alpha^{\phantom'}}$ & $\overline{C}_\alpha$ & 1  & -\,1 & 1\\\hline
$\mathcal{C}^{A^{\phantom'}}$ & $\overline{\mathcal{P}}_A$ & 2  & -\,2 & 0 &$\lambda^{A^{\phantom'}}$ & $\pi_A$ & 1  & -\,1 & 1\\ \hline
$\mathcal{C}^{{A_1}^{\phantom'}}$ & $\overline{\mathcal{P}}_{A_1}$ & 3  & -\,3 & 1 &$\mathcal{P}^{A^{\phantom'}}$ & $\overline{\mathcal{C}}_A$ & 2  & -\,2 & 0\\ \hline
$\lambda'^{{A_1}^{\phantom'}}$ & $\pi'{}_{A_1}$ & 1  & -\,1 & 1 &$\lambda^{{A_1}^{\phantom'}}$ & $\pi_{A_1}$ & 2  & -\,2 & 0\\ \hline
$\mathcal{P}'^{{A_1}^{\phantom'}}$ & $\overline{\mathcal{C}}{}'{}_{A_1}$ & 2  & -\,2 & 0 &$\mathcal{P}^{{A_1}^{\phantom'}}$ & $\overline{\mathcal{C}}_{A_1}$ & 3  & -\,3 & 1\\ \hline
\end{tabular}
\end{center}
\end{table}

\section{Example: Maxwell-like spin 2 model}

\noindent Let us consider a theory of symmetric traceful second-rank tensor in $4d$ Minkowski space with the metric $\eta_{\mu\nu}=\text{diag}(1,-1,-1,-1)$, described by the Maxwell-like Lagrangian \cite{Campoleoni:2012th},
\begin{equation}\label{MlLagr}
    \mathcal{L}=-\frac{1}{2}\partial_\rho h_{\mu\nu}\partial^\rho h_{\mu\nu}+\partial^\mu h_{\mu\nu}\partial_\rho h^{\rho\nu}\,,
\end{equation}
where $\mu, \nu, \rho=0,1,2,3$. The modified Noether identities (\ref{LNI}) read
\begin{equation}
    -\,2\partial^\mu\frac{\delta S}{\delta h^{\mu\nu}}+\partial_\nu\tau=0\,, \quad 
    \tau\equiv-\,2\partial_\mu\partial_\nu h^{\mu\nu}\,.
\end{equation}
A single completion function, $\mathcal{T}\equiv\tau-\Lambda$,
can be resolved with respect to $\Lambda$, $\Lambda\approx\tau$, i.e. the global conserved quantity $J$  coincides with $\tau$.

The unfree gauge symmetry transformations (\ref{Lunfree}) for the theory (\ref{MlLagr}) read
\begin{equation}
    \delta_\epsilon h^{\mu\nu}=
    \partial^\mu\epsilon^\nu+\partial^\nu\epsilon^\mu\,, \quad \partial_\mu\epsilon^\mu=0\,,
\end{equation}
while corresponding reducible transformations read
\begin{equation}
    \delta_\varepsilon h^{\mu\nu}=\partial^\mu\partial_\rho\varepsilon^{\rho\nu}+
    \partial^\nu\partial_\rho\varepsilon^{\rho\mu}\,, \,\,
    \delta_\omega \varepsilon^{\mu\nu}=\partial_\lambda\omega^{\lambda\mu\nu}\,, \,\, \delta_\eta\omega^{\mu\nu\lambda}=-\,\varepsilon^{\mu\nu\lambda\rho}\partial_\rho \eta\,,
\end{equation}
where $\varepsilon^{\mu\nu}, \omega^{\mu\nu\lambda}$ are totally antisymmetric tensors, $\varepsilon^{\mu\nu\lambda\rho}$ is the Levi-Civita symbol, $\epsilon^\mu=\partial_\nu\varepsilon^{\nu\mu}$.

Introducing the momenta
\begin{equation}
    \displaystyle \Pi_{ij}\equiv-\dot{h}_{ij}+\partial_i\lambda_j+\partial_j\lambda_i\,, \quad \Pi\equiv\dot{h}+2\partial_i\lambda^i\,,
\end{equation}
we arrive to the Hamiltonian action for the theory (\ref{MlLagr}),
\begin{equation}\label{MlSH}
    \begin{array}{c}
    \displaystyle S_H=\int d^4 x \big(\Pi_{ij}\dot{h}^{ij}+\Pi\dot{h}
    -H-\lambda^iT_i\big)\,, \quad T_i=2\big(\partial_i\Pi-\partial^j\Pi_{ji}\big)\,,
    \\[2mm]
    \displaystyle H=\frac{1}{2}\big(\Pi^2-\Pi_{ij}\Pi^{ij}\big)+\frac{1}{2}\big(\partial_ih_{jk}\partial^ih^{jk}+\partial_ih\partial^ih\big)-\partial^ih_{ij}\partial_kh^{kj}\,,
    \end{array}
\end{equation}
where $i,j,k=1,2,3$, $h^{00}\equiv h$, $h^{0i}\equiv\lambda^i$.

According to the Dirac-Bergmann algorithm, conservation of primary constraints $T_i$ leads to a single irreducible secondary constraint $\tau$,
\begin{equation}\label{MlHtau}
    \{T_i\,, H_T\}=-\,\partial_i \tau\,, \quad \tau=-\,2(\Delta h+\partial_i\partial_jh^{ij})-\Lambda\,, \quad \Lambda=const\,.
\end{equation}
In this case, we arrive to unfree symmetry transformations
\begin{equation}
    \delta_\epsilon h^{ij}=
    \partial^i\epsilon^j+\partial^j\epsilon^i\,, \,\, \delta_\epsilon h=-\,2\partial_i\epsilon^i\,, \,\, \delta_\epsilon \lambda^i=
    \dot{\epsilon}^i+\partial^i\epsilon\,, \,\, \dot{\epsilon}+\partial_i\epsilon^i=0\,,
\end{equation}
cf. (\ref{unfreetransf}). The corresponding BRST-charge reads
\begin{equation}\label{MlQ}
    Q=2\big(\partial_i\Pi-\partial^j\Pi_{ji}\big)-2\big(\Delta h+\partial_i\partial_jh^{ij}\big)+\pi_iP^i\,.
\end{equation}

If we consider $\widetilde{\tau}_i\equiv-\,\partial_i\tau$ as reducible secondary constraints, the corresponding gauge symmetry will be reducible, cf. (\ref{redtransf}),
\begin{equation}
    \begin{array}{c}
    \delta_\varepsilon h^{ij}
    =\partial^i(\dot{\varepsilon}^{j}+\partial_k\varepsilon^{kj})+
    \partial^j(\dot{\varepsilon}^{i}+\partial_k\varepsilon^{ki})\,, \quad \delta_\varepsilon h=
    -\,2\partial_i\dot{\varepsilon}^i\,,\\[2mm] 
    \delta_\varepsilon \lambda^i=
    \ddot{\varepsilon}^i+\partial_j\dot{\varepsilon}^{ji}-
    \partial^i\partial_j\varepsilon^j\,; \,\, 
    \delta_\omega\varepsilon^{ij}=\dot{\omega}^{ij}+\varepsilon^{ijk}\partial_k{\omega}\,, \,\, \delta_\omega\varepsilon^i=
    -\,\partial_j\omega^{ji}\,;\\[2mm]
    \delta_\eta\omega^{ij}=-\,\varepsilon^{ijk}\partial_k\eta\,, \quad
    \delta_\eta\omega=\dot{\eta}\,,
    \end{array}
\end{equation}
where $\varepsilon^{ij}, \omega^{ij}$ are totally antisymmetric tensors, and $\varepsilon^{ijk}\equiv\varepsilon^{0ijk}$ is the Levi-Civita symbol. 
The corresponding BRST charge (\ref{widetildeQ}) has the form
\begin{equation}\label{MlwidetildeQ}
    \begin{array}{c}
    \widetilde{Q}=2\big(\partial_i\Pi-\partial^j\Pi_{ji}
    \big)C^i+2\partial_i\big(\Delta h
    +\partial_j\partial_kh^{jk}\big)
    \mathcal{C}^i-\overline{\mathcal{P}}_i
    \partial_j\mathcal{C}^{ji}
    \\[2mm]
    -\,\overline{\mathcal{P}}_{ij}
    \varepsilon^{ijk}\partial_k\mathcal{C}
    +\pi_iP^i+\pi_{ij}\mathcal{P}^{ij}+\pi\mathcal{C}
    +\pi'\mathcal{P}'\,.
    \end{array}
\end{equation}

The theory (\ref{MlSH}) admits two different BRST complexes defined by (\ref{MlQ}) and (\ref{MlwidetildeQ}). The irreducible secondary constraint $\tau$ (\ref{MlHtau}) is BRST-exact with respect to $Q$ (\ref{MlQ}), $\tau=\{\overline{P},Q\}$, 
so the modular parameter $\Lambda$ is a trivial quantity, that corresponds to the case of fixed field asymptotics. At the same time, $\tau$ is BRST-closed, but not BRST-exact with respect to $\widetilde{Q}$ (\ref{MlwidetildeQ}),
\begin{equation}
    \{\tau,\widetilde{Q}\}=0\,, \quad \tau=\{\psi,\widetilde{Q}\}\,, \quad \psi=\overline{\mathcal{P}}_i\psi^i\,,
\end{equation}
as $\tau=\widetilde{\tau}_i\psi^i$, where $\psi^i$ is a non-local operator inverse to partial derivative. So, $\Lambda$ is a physical quantity, and various field asymptotics are admissible.

\section{Conclusion}
\noindent We demonstrated that the theories with unfree gauge symmetry can be self-consistently described by two different BRST complexes. One of them corresponds to the unfree gauge symmetry as such, while another one is connected with the alternative description of the same symmetry with reducible gauge transformations with unrestricted gauge parameters. These two descriptions are related, but not equivalent. The choice between them depends on the physical problem setting. For the first BRST complex, the global conserved quantities are BRST-exact, and for the second one, they are BRST-closed, but non-trivial.

\vspace{0.4 cm} \noindent \textbf{Acknowledgments.} The work is supported by the Foundation for the Advancement of Theoretical Physics and Mathematics ``BASIS".

\end{document}